\documentclass[12pt]{article}
\usepackage{jheppub}
\usepackage[T1]{fontenc}
\usepackage{amsmath,amssymb,amsfonts,graphicx,subfigure,
slashed,amsthm,mathtools,upgreek,enumerate,tensor}
\usepackage[dvipsnames]{xcolor}
\usepackage{arydshln}
\usepackage{braket}
\usepackage{color}
\usepackage{cases}
\usepackage{hyperref}
\usepackage[utf8]{inputenc}
\usepackage[titletoc]{appendix}

\definecolor{phthaloblue}{rgb}{0.0, 0.06, 0.54}
\definecolor{purple}{rgb}{0.5 ,0, 0.7}
\definecolor{bluegreen}{rgb}{0, 0.45, 0.35}
\definecolor{sakura}{rgb}{1 ,0.52, 0.74}
\definecolor{wakakusa}{rgb}{0.45 ,0.74, 0}
\definecolor{brown}{rgb}{0.48 ,0.23, 0}
\definecolor{skyblue}{rgb}{0.21 ,0.7, 1.}
\definecolor{purplegray}{rgb}{0.35,0.35,0.73}
\usepackage{hyperref}
\hypersetup{colorlinks=true, linkcolor=bluegreen, citecolor=purplegray, urlcolor=purplegray}

\usepackage{acro}
\DeclareAcronym{gr}{
    short=GR ,
    long=general relativity
}
\DeclareAcronym{bfv}{
    short=BFV ,
    long=Batalin-Fradkin-Vilkovisky
}
\DeclareAcronym{adm}{
    short=ADM ,
    long={Arnowitt, Deser and Misner}
}

\title{\bf Summing over Non-singular Paths in Quantum 
Cosmology}

\author{\large Hiroki Matsui${}^a$}

\emailAdd{hiroki.matsui@yukawa.kyoto-u.ac.jp}

\affiliation{${}^a$Center for Gravitational Physics and Quantum Information,
Yukawa Institute for Theoretical Physics, Kyoto University,
Kitashirakawa Oiwakecho, Sakyo-ku,
Kyoto 606-8502, JAPAN}

\abstract{In this paper we provide the DeWitt propagator and its wave function in quantum cosmology using the path integral formulation of quantum gravity. The DeWitt boundary condition is introduced as a way to avoid the Big Bang singularity by positing that the wave function of the universe vanishes near the Big Bang.
However, there is currently no clear definition of the DeWitt boundary condition in the path integral formulation. To address this issue, we use the image method, which eliminates singular paths in the forbidden region of the infinite potential, and apply this method to quantum cosmology based on the 
Batalin-Fradkin-Vilkovisky formulation of the path integral. 
We investigate the validity of the image method, and in particular, find that this method is only appropriate when the potential has symmetry with respect to the boundary. We then show that the DeWitt propagator and the DeWitt wave function derived from the image method are consistent with solutions of the Wheeler-DeWitt equation for certain models of quantum cosmology. }

\preprint{YITP-23-55}
\begin{document}
\maketitle

\section{Introduction}

One of the most intriguing questions in modern cosmology is how the primordial universe emerged from the Planck era.  The standard theory describing gravity, \ac{gr}, is a classical theory and is unable to describe the creation and evolution of a Planck-sized primordial universe in which the effects of quantum gravity become dominant. Quantum cosmology, on the other hand, aims to address this question by utilizing the theory of quantum gravity and introducing the concept of the wave function of the universe, which is represented by the wave-functional of quantum gravity, $\Psi[g^{(3)},\phi]$, where $g^{(3)}$ is the spatial metric induced on a 3-geometry and $\phi$ is the matter field configuration.
In this framework, the wave function of the universe is given by a particular solution of the Wheeler-DeWitt equation~\cite{DeWitt:1967yk} satisfying certain boundary conditions. The Wheeler-DeWitt equation is derived from the Hamiltonian formulation of gravity by imposing the constraints of diffeomorphism invariance, and describes the quantum dynamics of gravity in terms of the canonical variables of gravity. 
Alternatively, the wave function of the universe can be formulated by the path integral of quantum gravity, $\Psi[g^{(3)},\phi]=\int\mathcal{D} g^{(4)}\mathcal{D}\phi\, e^{i S[g^{(4)},\phi]/\hbar}$, where the 4-dimensional metric $g^{(4)}$ is restricted to those which induce $g^{(3)}$ on the 3-geometry~\cite{Halliwell:1988wc}.

In quantum cosmology, boundary conditions on the wave function of the universe are usually motivated by physical considerations or mathematical consistency. For example, one can impose the no-boundary condition~\cite{Hawking:1981gb, Hartle:1983ai, Hawking:1983hj}, which states that the wave function of the universe is given by a sum over all compact and regular geometries. This implies that the universe has no initial singularity or boundary, and that the cosmological history is uniquely determined by its wave function. 
One can also impose the tunneling condition~\cite{Vilenkin:1984wp, Vilenkin:1986cy, Vilenkin:1987kf}, which states that the universe emerged from no space-time, i.e. from nothing, by quantum tunneling, and predicts the initial state of the inflationary universe. 
\footnote{
The related early proposals can be found in the literature~\cite{Linde:1983cm,Linde:1983mx,Linde:1984ir,Rubakov:1984bh,Zeldovich:1984vk}.}

These boundary conditions are expected to select a unique solution for the wave function of the universe and provide a probabilistic interpretation for its quantum state. 
As solutions in the Wheeler-DeWitt equation, one can easily find its wave function for the no-boundary or tunneling proposals in the minisuperspace approximation~\cite{Vilenkin:1984wp, Vilenkin:1986cy, Vilenkin:1987kf}. 
On the other hand, the formulation of the no-boundary and tunneling proposals based on path integrals had not been explicitly formulated due to various problems with quantum gravity. Recently, however, these formulations based on the Lorentzian path integral have been proposed in the literature~\cite{Feldbrugge:2017kzv}, using the \ac{bfv} formulation of the path integral and Picard-Lefschetz theory.
Also based on this formulation, small perturbations around the background are found to be inverse Gaussian and uncontrollable~\cite{Feldbrugge:2017fcc, Feldbrugge:2017mbc}. This perturbation problem has recently been discussed in the literature~\cite{Feldbrugge:2017fcc, DiazDorronsoro:2017hti, Feldbrugge:2017mbc, Feldbrugge:2018gin, DiazDorronsoro:2018wro,Halliwell:2018ejl,Janssen:2019sex, Vilenkin:2018dch, Vilenkin:2018oja, Bojowald:2018gdt, DiTucci:2018fdg, DiTucci:2019dji, DiTucci:2019bui, Lehners:2021jmv,Matsui:2022lfj}.
In quantum cosmology, these no-boundary or tunneling proposals are widely studied.

On the other hand, there is another boundary condition such that the wave function of the universe vanishes near the Big Bang singularity in the Wheeler-DeWitt equation. This is called the {\it DeWitt boundary condition} and was first proposed as a boundary condition in quantum cosmology~\cite{DeWitt:1967yk}. The DeWitt boundary condition states $\Psi[g^{(3)},\phi]=0$ for all singular 3-metric of spacial hypersurfaces $g^{(3)}$ and sufficiently removes or avoids the Big Bang singularity.
\footnote{The role of the DeWitt boundary condition in resolving singularities continues to be a subject of debate although this is widely discussed as a solution to the singularity problem in quantum gravity~\cite{Hajicek:2001yd,Dabrowski:2006dd,Kiefer:2019csi,Kiefer:2019bxk}. Some argue that the DeWitt boundary condition may not contribute significantly, proposing that the vacuum expectation values of any operator for a quantum state $|\Psi\rangle$ should vanish to avoid the singularities~\cite{Lund:1973zz,Gotay:1980xk,Gotay:1983kvm,Jalalzadeh:2022dlj}. 
}
There has been some work on the DeWitt proposal, e.g. Refs~\cite{Jalalzadeh:2014jka,Jalalzadeh:2014zaa,Rostami:2015kua,Jalalzadeh:2016gqs}. Although there is no DeWitt wave function that behaves perturbatively well when inhomogeneous tensor perturbations are taken into account in \ac{gr}, the unstable perturbation problem can be solved in Ho\v{r}ava-Lifshitz gravity~\cite{Matsui:2021yte,Martens:2022dtd}. On the other hand, much work on the DeWitt proposal has been done only on the Hamiltonian formulation, i.e. the Wheeler-DeWitt equation, there is no clear definition of the DeWitt proposal based on the path integral formulation of quantum gravity~\cite{Hawking:1978jz}.
\footnote{Following the Hartle-Hawking proposal, it has been shown that the one-loop wave function based on the path integral of gravity satisfies the DeWitt proposal with the appropriate boundary conditions on the perturbations~\cite{Esposito:2023ymk}. On the other hand, the DeWitt wave functions introduced in this paper are established at the background level.}
This is one of the main topics of this paper.

In this paper, we present the DeWitt wave function in the path integral formulation of quantum gravity. First of all, imposing boundary conditions on the wave functions based on the path integral formulation is known to be conceptually and technically unclear even in quantum mechanics, and the infinite potential is somehow troublesome in the path integral formulation~\cite{feynman2010quantum}.  The simplest method to eliminate the paths in the forbidden region of this infinite potential is the image method proposed by Refs~\cite{Janke:1979fv,1981AmJPh..49..843G}. This method has been applied to the infinite (square) well potential and it has been shown that the derived propagator agrees with the path integral formulation expanded by the eigenfunctions. The validity of the image method has been investigated in Refs.~\cite{1984PhLA..106..212S,1985NuPhB.257..799A,
1993PhRvA..48.3445N,1997JPhA...30.5993A,2011arXiv1112.3674D}. However, we will show that the image method is only applicable for the symmetric potential with respect to the boundary in Appendix~\ref{sec:appendix1}.   For asymmetric potentials, the restricted propagator of the image method does not give the correct wave function that satisfies the Schr\"{o}dinger equation. Based on this consideration, we will apply this image method to the framework of quantum cosmology using the \ac{bfv} formulation of the path integral~\cite{Fradkin:1975cq,Batalin:1977pb}, and provide the DeWitt propagator and the DeWitt wave function based on the Lorentzian path integral formulation and the image method. For the symmetric potential written by the scale factor with respect to the boundary, we show that these are consistent with the solutions of the Wheeler-DeWitt equation, and suggest the singularity-free origin of the universe.

This paper is organized as follows: in Section~\ref{sec:image-method} we briefly review the image method in quantum mechanics. We discuss several examples such as a free particle in an infinite potential barrier, an infinite square well potential, and a half-harmonic oscillator model. In these examples, the image method can correctly provide the restricted propagator in quantum mechanics.  In Section~\ref{sec:bfv-formalism} we briefly review the path integral formulation of quantum gravity using the \ac{bfv} formulation of the gravitational path integral and Picard-Lefschetz theory. It is shown that the Lorentzian path integral formulation elegantly derives the wave function of the universe, which correctly satisfies the Wheeler-DeWitt equation. In Section~\ref{sec:DeWitt-propagator} we define the DeWitt propagator and the DeWitt wave function based on the Lorentzian path integral formulation and the image method. We show that such DeWitt propagator and wave function are consistent with the solutions of the Wheeler-DeWitt equation. Finally, we conclude our work in Section~\ref{sec:conclusion}.

\section{Restricted propagator in quantum mechanics}
\label{sec:image-method}

In this section, we will explore the propagator for a free particle in an infinite potential barrier and review the image method in the path integral formulation of quantum mechanics. The image method is a common mathematical technique for solving differential equations where a solution satisfying certain boundary conditions can be easily derived by adding a mirror image to the symmetry boundary of the domain. The image method in path integrals is defined as follows~\cite{1981AmJPh..49..843G}: 
the restricted propagator $K_R\left[x', t'; x,t\right]$, which satisfies the boundary conditions such as the infinite potential barrier, is given by 
the difference of the particle propagators starting 
from the point $(x',t)$ and its image point $(-x',t)$.

The propagator $K_R\left[x', t'; x,t\right]$ in the restricted domain, $0 < x < \infty$ is written as,
\begin{align}\label{restricted-propagator}
K_R\left[x', t'; x,t\right] :=  K\left[x', t'; x,t\right] 
- K\left[-x', t';x,t\right] \; ,
\end{align}
where $K\left[x', t'; x,t\right]$ is the usual propagator and 
satisfies the Schr\"{o}dinger equation for the particle with mass $m$, 
\begin{equation}\label{Schrodinger-equation}
\left[
-\frac{\hbar^2}{2m}\frac{\partial^2}{\partial x'^2}+V(x')-
i\hbar \frac{\partial}{\partial t'}\right]K\left[x', t'; x,t\right]=-i\hbar\delta(x'-x)\delta(t'-t)\,.
\end{equation}
The wave function for the restricted propagator~\eqref{restricted-propagator}  is given by
\begin{align}
\Psi\left[x', t'\right] =\int\Psi\left[x, t\right]K_R\left[x', t'; x,t\right] 
\mathrm{d}x \;,
\end{align}
which satisfies the boundary condition $\Psi\left[0, t'\right]= 0$
\footnote{On the other hand, there is another definition of the restricted propagator $K_R\left[x', t'; x,t\right]$~\cite{1981AmJPh..49..843G} by considering 
the different propagators originating 
from the point $(x,t)$ and its image point $(-x,t)$, and this is given by $K_R\left[x', t'; x,t\right] :=  K\left[x', t'; x,t\right] 
- K\left[x', t';-x,t\right]$
which does not lead to $\Psi\left[0, t'\right]= 0$.
Thus, we do not consider the propagator in this paper.}.
We note that this definition does not automatically satisfy the Schr\"{o}dinger equation~\eqref{Schrodinger-equation}, but in symmetric potentials with $V(x')=V(-x')$, the propagator~\eqref{restricted-propagator}  
derives an exact restricted propagator. 
We provide proof in Appendix~\ref{sec:appendix1} and the wave function derived from the restricted propagator $K_R\left[x', t'; x,t\right]$ satisfies the Schr\"{o}dinger equation when the potential exhibits symmetry with respect to the boundary.

First, we consider the propagator for a free particle in the infinite potential barrier,
$0< x<\infty$ as an example of the image method in quantum mechanics. 
The Hamiltonian for this system can be written as $\hat{H} = -\frac{\hbar^2}{2m}\frac{\partial^2}{\partial x^2}+V(x)$,
where $V(x)$ is given by
\begin{align}
V(x)=\left\{
\begin{array}{l l}
\infty &\quad \text{for $x \le0$}\\
0 &\quad \text{for $x > 0$}\\
\end{array}\right.~.
\end{align}
The corresponding energy eigenfunctions for the system are
\begin{align}
\Psi_k= \left\{
\begin{array}{l l}
\ 0  &\quad \text{for $x \le0$}\\
\sqrt{\frac{2m}{\hbar^2\pi k}}\, \sin kx &\quad \text{for $x > 0$}\\
\end{array}\right.~,
\end{align}
which varnishes at $x=0$ and the wave-number $k$ is 
related to the energy, $E=\frac{\hbar^2k^2}{2m}$.
By using the above eigenfunctions we can derive the propagator in the infinite potential barrier,
\begin{align}\label{e-propagator}
\begin{split}
K_R\left[x', t'; x,t\right] &= \int \mathrm{d}E\, \Psi_k^*[x]\Psi_k[x']e^{-{iE(t'-t)\over\hbar}}\\
&=\int^{\infty}_0 \frac{2\mathrm{d}k}{\pi}\sin kx\sin kx'e^{-\frac{i\hbar k^2}{2m}(t'-t)}.
\end{split}
\end{align}

Hereafter, we shall show that the restricted propagator~\eqref{restricted-propagator} 
derived by the image method is consistent with the above propagator~\eqref{e-propagator}
in the infinite potential barrier. In the definition of the restricted propagator~\eqref{restricted-propagator} 
we have 
\begin{align}\label{rb-propagator}
K_R\left[x', t'; x,t\right]&= K_F\left[x', t'; x,t\right] - K_F\left[-x', t'; x,t\right]\notag \\
&=\left(\frac{m}{2i\pi\hbar(t'-t)}\right)^{1/2} \left[e^{im(x'-x)^2/2\hbar(t'-t)}-e^{im(x'+x)^2/2\hbar(t'-t)}\right],
\end{align}
where $K_F\left[x', t'; x,t\right]$ represents the propagator for the free particle in the full space
and can be written as the Gaussian integration,
\begin{align}\label{free-propagator}
K_F\left[x', t'; x,t\right] &=\left(\frac{m}{2i\pi\hbar(t'-t)}\right)^{1/2} e^{im(x'-x)^2/2\hbar(t'-t)}\notag \\
&=\int^{\infty}_{-\infty} \frac{\mathrm{d}k}{2\pi}\exp
\left[ik(x'-x)-i\frac{\hbar k^2}{2m}(t'-t)\right] .
\end{align}
By inserting this expression~\eqref{free-propagator} to Eq. (\ref{rb-propagator})
we have
\begin{align}
&K_R\left[x', t'; x,t\right] = K_F\left[x', t'; x,t\right] - K_F\left[-x', t';x,t\right]\notag \\
&=\int^{\infty}_{-\infty} \frac{\mathrm{d}k}{2\pi}\exp
\left[-i\frac{\hbar k^2}{2m}(t'-t)\right] 
\left[\exp\left[ ik(x'-x)\right]-\exp\left[ ik(x'+x)\right]\right]\notag \\
&=-\int^{\infty}_{0} \frac{\mathrm{d}k}{2\pi}\exp
\left[-i\frac{\hbar k^2}{2m}(t'-t)\right] 
\left[\exp\left[ ikx'\right]-\exp\left[ -ikx'\right]\right]
\left[\exp\left[ ikx\right]-\exp\left[ -ikx\right]\right]\notag \\
&=\int^{\infty}_{0} \frac{2\mathrm{d}k}{\pi}\sin kx\sin kx'\exp
\left[-i\frac{\hbar k^2}{2m}(t'-t)\right] ,
\end{align}
which is consistent with the propagator~\eqref{e-propagator} expanded by the eigenfunctions
in the infinite potential barrier.

On the other hand, the image method can be applied to the infinite square well 
potential and it is shown that $K_R\left[x', t'; x,t\right] $ agrees with the propagator
expanded by the eigenfunctions.
In infinite square well potential, the particle repeatedly reflects 
between potentials so that the elimination of forbidden paths is also non-trivial.
The propagator $K_R\left[x', t'; x,t\right]$ 
in the infinite square well, $0 < x < L$ is given by
\begin{align}\label{square-well-propagator}
K_L\left[x', t'; x,t\right] = \sum_{n=-\infty}^{\infty}K_F[x'+2nL,t';x,t] - \sum_{n=-\infty}^{\infty}K_F[-x'+2nL,t';x,t]\; ,
\end{align}
which is consistent with the correct propagator
expanded by the eigenfunctions~\cite{Janke:1979fv,1981AmJPh..49..843G}. Note that in the limit $L\to\infty$, 
the propagator $K_L\left[x', t'; x,t\right]$ is identical with
$K_R\left[x', t'; x,t\right]$, and we have, 
\begin{align}
\begin{split}
K_L\left[x', t'; x,t\right] &= \sum_{n=-\infty}^{\infty}K_F[x'+2n\cdot\infty,t';x,t] 
- \sum_{n=-\infty}^{\infty}K_F[-x'+2n\cdot\infty,t';x,t]\\
&= K_F[x',t';x,t] - K_F[-x',t';x,t]=K_R\left[x', t'; x,t\right] .
\end{split}
\end{align}

Next, we will discuss the particle propagator in the half-harmonic oscillator potential
where the particle is limited in $0< x<\infty$.
This discussion was developed by the literature~\cite{2011arXiv1112.3674D}, but we will briefly discuss the results to confirm for the reader the validity of the restricted propagator~\eqref{restricted-propagator}.
The half-harmonic oscillator potential is given by
\begin{align}
V(x)=\left\{
\begin{array}{l l}
\infty &\quad \text{for $x \le0$}\\
\frac{m\omega^2}{2} x^2 &\quad \text{for $x > 0$}\\
\end{array}\right.~.
\end{align}
We start with the propagator $K_{H}\left[x', t'; x,t\right]$ 
for a harmonic oscillator in $-\infty<x<\infty$~\cite{feynman2010quantum},
\begin{align}
K_{H}\left[x', t'; x,t\right]= \sqrt{\frac{m\omega}{2i\pi\hbar\sin\omega T}}~
\exp{\left(\frac{im\omega}{2\hbar\,\sin\omega T}\left[\left(x'^2+x^2\right)\,\cos\omega T-2x'x
\right]\right)},
\end{align}
where $T=(t'-t)$. By using the image method, the restricted propagator is given by
\begin{align}
&K_{R}\left[x', t'; x,t\right]
=K_{H}\left[x', t'; x,t\right] - K_{H}\left[-x', t';x,t\right]\\
&=\sqrt{\frac{m\omega}{2\pi\hbar\,i\,\sin\omega T}}~
\exp{\left(\frac{im\omega}{2\hbar\,\sin\omega T}
\left(x'^2+x^2\right)\,\cos\omega T
\right)}
\left[\exp{\left(-\,\frac{im\omega x'x}{\hbar\sin\omega T}\right)}
-\exp{\left(\frac{im\omega x'x}{\hbar\sin\omega T}\right)}
\right].\nonumber
\end{align}

In order to verify the above restricted propagator,  
we utilize the propagator expanded by the eigenfunctions,
\begin{align}
K\left[x', t'; x,t\right] =
\sum_{n=0}^\infty \Psi_n^*[x]\Psi_n[x']e^{-\frac{i}{\hbar}E_nT}~.
\end{align}
Taking $x=x'$, we have the following relation for the eigenvalues,
\begin{align}\label{eigenvalue-relation}
\int_0^\infty\mathrm{d}x\, K\left[x', t'; x,t\right]=
\sum_{n=0}^\infty e^{-\frac{i}{\hbar}E_n\tau} 
\int_0^\infty\mathrm{d}x\left|\Psi_n[x]\right|^2=
\sum_{n=0}^\infty e^{-\frac{i}{\hbar}E_nT} ~,
\end{align}
where we used the orthonormality of $\Psi_n[x]$. 
Following Eq.~\eqref{eigenvalue-relation}, 
we integrate $K_{R}\left[x', t'; x,t\right]$,
\begin{align}
\int_0^\infty\mathrm{d}x\, K\left[x', t'; x,t\right]=\sum_{n=0}^\infty
e^{-i\left(2n+\frac{3}{2} \right)\omega T},
\end{align}
and we have the correct eigenvalues for the half-harmonic oscillator,
\begin{align}
        E_n = \left(2n+\frac{3}{2}\right) \hbar\omega\,, ~n=0,1,2,3,\cdots .
\end{align}
which suggests that the propagator $K_R[x', t'; x,t]$ derived by the image method
reproduces correct eigenvalues for the half-harmonic oscillator.

We have shown that the image method is justified as the correct path integral method for simple infinite potential problems.
However, its applicability to general potentials remains less clear. 
The validity of the image method has been investigated in the literature~\cite{1984PhLA..106..212S,1985NuPhB.257..799A,
1993PhRvA..48.3445N,1997JPhA...30.5993A,2011arXiv1112.3674D}. In practice, however, it is unclear whether the image method can introduce the correct propagator for asymmetric potentials to the boundary conditions. The wave function defined by the restricted propagator with asymmetric potentials does not satisfy the Schr\"{o}dinger equation~\eqref{Schrodinger-equation}, as shown in Appendix~\ref{sec:appendix1}. In fact, from a mathematical perspective, the image method is not applicable to the asymmetric potential. Thus, in Section~\ref{sec:DeWitt-propagator} we will consider the application of the image method to quantum cosmology where the potential of the scale factor is symmetric with respect to the singularity.

\section{Lorentzian path integral in 
Batalin-Fradkin-Vilkovisky formalism}
\label{sec:bfv-formalism}

Before starting our original discussions, we provide a review of the path integral formulation of quantum gravity. We specifically focus on the conventional approaches to path integral methods within minisuperspace models, 
employing the \ac{bfv} formulation and the Picard-Lefschetz theory.
The gravitational transition amplitude from a three-geometry $g_{0}$ on an initial spacelike surface to a three-geometry $g_{1}$ on a final spacelike surface where we drop the three for $g_{0,1}^{(3)}$, is given by
\begin{equation}\label{G-amplitude1}
G[g_{1};g_{0}]:=  \int_{\cal M} \mathcal{D}g_{\mu\nu}
~ \exp \left({i S[g_{\mu\nu}]}/{\hbar}\right) ~,
\end{equation}
where $S[g_{\mu\nu}]$ is the Einstein-Hilbert action in \ac{gr}. 
The Einstein-Hilbert action 
with a positive cosmological constant $\Lambda$ and a boundary term 
is written as
\begin{equation}
S[g_{\mu\nu}] = \frac{1}{2}\int_{\cal M} \mathrm{d}^4x \sqrt{-g} \left( R - 2 \Lambda\right) + \int_{\cal \partial M} \mathrm{d}^3y \sqrt{g^{(3)}} \mathcal{K}\,,
\end{equation} 
where we set $8\pi G = 1$.
The second term is the Gibbons-Hawking-York boundary term 
with the 3-metric $g^{(3)}_{ij}$ and the trace of the extrinsic
curvature $\mathcal{K}$ of the boundary $\partial {\cal M}$.
By using the minisuperspace approximation, the line element can be expressed as~\cite{Louko:1987wq},
\begin{align}\label{metric}
\textrm{d} s^2 &= \sigma^2\left( 
-\frac{N(t)^2}{q(t)}\textrm{d} t^2 + q(t)\left( \frac{ \text{d}r^2 }{1 -K r^2} + r^2 (\text{d}\theta^2 + \sin^2 \theta \text{d}\phi^2) \right) \right),
\end{align}
where $\sigma^2=2G/3\pi $, $q(t) \equiv a(t)^2$, $a(t)$ is the scale factor and $N(t)$ is the lapse function.
For simplicity, we assume the positive spatial curvature $K=1$.

We use the \ac{bfv} formalism~\cite{Fradkin:1975cq,Batalin:1977pb}
to perform the path integral of minisuperspace quantum gravity with reparametrization invariance, in which we perform an explicit integration over lapse and shift functions to ensure 
the Hamiltonian constraint. 
For the gauge-fixing choice $\dot{N}= 0$,
the \ac{bfv} path integral reads~\cite{Halliwell:1988wc},
\begin{equation}\label{BFV-path}
G[q_1;q_0] :=  \int \mathcal{D}q\mathcal{D}p_q\mathcal{D}\Pi\mathcal{D}N
\mathcal{D}\rho\mathcal{D}\bar{c}\mathcal{D}\bar\rho
\mathcal{D}c\,\exp(iS_{\rm BRS}/\hbar),
\end{equation}
where $p_q$ is conjugate momentum of $q$ and 
$S_{\rm BRS}$ is the Becchi-Rouet-Stora (BRS) invariant action which is given by 
\begin{equation}
S_{\rm BRS}:= \int_{t=0}^{t=1} {\rm d}t\left(p_q \dot{q} - N \mathcal{H} + \Pi \dot{N} +\bar\rho\dot{c} +\bar{c}\dot\rho-\bar\rho\rho\right),
\end{equation}
where $\mathcal{H}[q,p_q]$ is the Hamiltonian constraint,
$\mathcal{H}[q,p_q]=\frac{1}{2}\left(-4p_q^2+\lambda q-1 \right)
$ with $\lambda=\frac{\sigma^2}{3}\Lambda$
and $\Pi$ is a Lagrange multiplier and $\rho,\bar{c},\bar\rho$ and $c$ are ghost fields 
preserving the BRS symmetry under the transformations,
\begin{align}
\begin{split}
\delta q=\zeta c\frac{\partial \mathcal{H} }{\partial p_q}\,,\;\;\delta p_q=-\zeta c\frac{\partial \mathcal{H} }{\partial q}\,,\;\;\delta N= \zeta\rho\,,\\ \delta\bar{c}=-\zeta\Pi\,,\;\;\delta\bar\rho=-\zeta \mathcal{H} \,,\;\; \delta \Pi=\delta c=\delta \rho=0,
\end{split}
\end{align}
where $\zeta$ is a parameter.
The ghost and multiplier parts can be integrated, and eventually, we obtain
the following gravitational transition amplitude~\cite{Halliwell:1988wc}, 
\begin{align}\label{G-amplitude}
G[q_1;q_0] &= \int {\rm d}N \int\mathcal{D}q\,\mathcal{D}p_q \,\exp\left(i\int_{t=0}^{t=1} {\rm d}t\left(p_q \dot{q} - N \mathcal{H}\right)/ \hbar\right)\nonumber
\\& =  \int {\rm d}Ne^{iN/2} \int\mathcal{D}q\,\mathcal{D}p_q \,\exp\left(i\int_{\tau=0}^{\tau=N} {\rm d}\tau \left(p_q \dot{q} - \frac{1}{2}\left(-4p_q^2+\lambda q\right)\right)/ \hbar\right)\,,
\end{align}
where we redefine the time $\tau=N t$. The path integral~\eqref{G-amplitude} provides either solution to the Wheeler-DeWitt equation or Green's function depending on the choice of the integration domains of the lapse function $N$. 
One of the formal concerns with the formulation of the \ac{bfv} path integral~\eqref{G-amplitude} is that the scale factor $q$, is positive, but by definition, the path integral performs integration in $-\infty<q<\infty$. However, it should be emphasized that if the \ac{bfv} path integral is simply used to generate solutions to the Wheeler-DeWitt equation, there is no mathematical contradiction with $q\geq 0$~\cite{Halliwell:1988wc}. In this sense, applying the image method of performing path integral in negative region $q<0$ to the \ac{bfv} path integral~\eqref{G-amplitude} is not a mathematical contradiction.

First, let us evaluate the path integral expression~\eqref{G-amplitude} and take $\hbar=1$. 
Although the functional integration over $p_q$ and $q$ results in the standard quantum mechanical propagator 
for a system with a linear potential, it's rather difficult to perform the integration. Thus, we shall perform the integration in a different order and proceed with the functional integral over $p_q$ and $q$ as follows~\cite{Halliwell:1988wc},
\begin{align}
\begin{split}
&\int\mathcal{D}q\,\mathcal{D}p_q \,\exp\left(i\int_{\tau=0}^{\tau=N} {\rm d}\tau \left(p_q \dot{q} - \frac{1}{2}\left(-4p_q^2+\lambda q\right)\right)\right) \\
&=(2 \pi)^{-n} \int_{-\infty}^{\infty} {\rm d}p_{1 / 2} \cdots {\rm d}p_{n+1 / 2} \int_{-\infty}^{\infty} {\rm d}q_1 \cdots {\rm d}q_n \\
&\quad \times \exp \left[i \sum_{k=0}^n\left[p_{k+1 / 2}\left(q_{k+1}-q_k\right)-\frac{\epsilon}{2}\left(-4 p_{k+1 / 2}^2+\lambda q_k\right)\right]\right], 
\end{split}
\end{align}
where we take $\epsilon=N/(n+1)$ and $q_{n+1}=q_{1}$. 
Performing the $q$ integration, we obtain the following expression, 
\begin{align}
G[q_1;q_0]=\int_{-\infty}^{\infty} {\rm d}N \int_{-\infty}^{\infty} {\rm d}p_q \exp \left[i\left[\left(2 p_q^2+\frac{1-\lambda q_{1}}{2}\right)N-\lambda p_q N^2+\frac{\lambda^2 N^3}{6}+p_q\left(q_{1}-q_{0}\right)\right]\right].
\end{align}
The $p_q$ integration can be carried out to yield the standard result for the propagator with a linear potential,
but it is more efficient to first integrate $N$. By redefining the lapse function as $M=N-2p_q/\lambda$, 
we have the form of the two independent integrals~\cite{Halliwell:1988wc},
\begin{align}\label{eq:Halliwell-solutions}
G[q_1;q_0]&=
\int_{-\infty}^{\infty} {\rm d}M \int_{-\infty}^{\infty} {\rm d}p_q 
\exp \left(\frac{i \lambda^2}{6}M^3+\frac{i\left(1-\lambda q_1\right)}{2}M \right) \exp \left(\frac{i 4 p_q^3}{3 \lambda}+\frac{i p_q\left(1-\lambda q_0\right)}{\lambda}\right)\notag  \\
&=\frac{2 \pi^2}{(4 \lambda)^{1 / 3}} \mathrm{Ai}\left(\frac{1-\lambda q_1}{(2 \lambda)^{2 / 3}}\right) \mathrm{Ai}\left(\frac{1-\lambda q_0}{(2 \lambda)^{2 / 3}}\right) \,,
\end{align}
where $\mathrm{Ai}(x)$ is the Airy function. 
This correctly satisfies the Wheeler-DeWitt equation,
\begin{equation}\label{eq:WdW-equation}
\hat{\mathcal{H}} G[q]=\frac{N}{2}\left(4 \frac{d^2}{d q^2}+\lambda q-1\right) G[q]=0 \,.
\end{equation}
However, the solution~\eqref{eq:Halliwell-solutions} is only one of the linearly independent solutions of the Wheeler-DeWitt equation. A different independent solution can be obtained by changing the integral contour and, the contour running from $-\infty$ to $0$ and then from $0$ to $-i\infty$ yields a different wave function~\cite{Halliwell:1988wc}. In principle, 
choosing which integration contours is not obvious in the path integral~\eqref{G-amplitude}, and one must deal
with highly oscillatory integrals when the integration is performed and is carried out along the real axis. Fortunately, this problem can be surmounted with Picard-Lefschetz theory~\cite{Witten:2010cx}.
The Picard-Lefschetz theory complexifies the variables themselves and provides a unique way to find a complex integration contour
along the steepest descent paths which is called Lefschetz thimbles $\cal J_\sigma$. This theory proceed with such oscillatory integral as, 
\begin{equation}
\int_\mathcal{R} \mathrm{d}x \, \exp\left(i S[x]\right)
=\sum_\sigma n_\sigma \int_{\cal J_\sigma} \mathrm{d}x
\exp\left(i S[x]\right).
\end{equation} 
where $n_\sigma$ is the intersection number $\braket{\cal K_\sigma,\mathcal{R}}$
between the steepest ascent path $\cal K_\sigma$ and the 
original contour $\mathcal{R} $. 
Although unproven, the Picard-Lefschetz theory can be expected to be 
applicable to the path integral~\cite{Witten:2010cx}. 
By utilizing the Picard-Lefschetz theory for the path integral of gravity,
we have
\begin{equation}
G[g_{1};g_{0}]=\sum_\sigma n_\sigma \int_{\cal J_\sigma}\mathcal{D}g_{\mu\nu}
\exp\left(i S[g_{\mu\nu}]\right) ~.
\end{equation}
In the \ac{bfv} path integral~\eqref{G-amplitude}, the functional integration of the lapse function $N$ reduces the ordinary integration of $N$ and we can apply the Picard-Lefschetz theory to this in a rigorous way without the mathematical ambiguity. It was first suggested in Refs~\cite{Halliwell:1988ik,Halliwell:1989vu} to find contours in the complex lapse plane where this integral converges and Ref.~\cite{Feldbrugge:2017kzv} 
provides a Picard-Lefschetz formulation of the \ac{bfv} path integral
to perform such a complex lapse integral, 
\begin{equation}
G[q_{1};q_{0}]=\sum_\sigma n_\sigma \int_{\cal J_\sigma}\mathrm{d}N
\exp\left(i S_{\textrm{on-shell}}[q_{1},q_{0},N]\right) ~,
\end{equation}
where $S_{\textrm{on-shell}}[q_{1},q_{0},N]$ is the on-shell action, and we have proceeded the functional integration of the scale factor $q(t)$ by substituting the solution with the Dirichlet boundary condition $q(1)=q_{1}$ and $q(0)=q_{0}$ for the action and integrating the path integral of $q(t)$ over the Gaussian fluctuation around the solution.
We can also impose the Neumann or Robin boundary conditions on the solutions and perform the path integral for those boundary conditions~\cite{DiTucci:2019dji,DiTucci:2019bui}. Still, we consider only the above Dirichlet boundary condition for simplicity. We provide a review of the Picard-Lefschetz formulation in the Appendix~\ref{sec:appendix2}.
Below, by using the \ac{bfv} path integral, 
and this Picard-Lefschetz formulation, we discuss the restricted propagator 
in quantum cosmology.

\section{DeWitt wave function in Lorentzian quantum cosmology}
\label{sec:DeWitt-propagator}

In this section, we will apply the image method to Lorentzian quantum cosmology on the basis of the \ac{bfv} path integral.
In quantum cosmology, there is a proposal to solve the initial singularity problem called the DeWitt boundary condition~\cite{DeWitt:1967yk}. This suggests that the wave function vanishes at the initial singularity, and can be defined by imposing this boundary condition in the Wheeler-DeWitt equation. We will consider how the corresponding propagator to satisfy the DeWitt boundary proposal can be constructed. As shown previously, the image method provides the restricted propagator in quantum mechanics which leads to the vanishing wave function in the boundary, and we apply this method in quantum cosmology.

We propose the restricted cosmological propagator which
is defined in the region 
$0 < q < \infty$,
\begin{equation}\label{DW-propagator}
G_{\rm DW}[q_1;q_0] :=  G[q_1;q_0]-G[-q_1;q_0]\,,
\end{equation}
which indeed leads to the DeWitt boundary condition, $G_{\rm DW}[0;q_0]=0$.
We note that integrating the lapse function $N$ from $0$ to $\infty$ or from $-\infty$ to $\infty$ in the \ac{bfv} path integral yields the propagator or wave function.
Therefore, we shall call this restricted 
propagator~\eqref{DW-propagator} DeWitt propagator or DeWitt wave function.
\footnote{
The image propagator should be defined as 
$G[-q_1;q_0]:=G[q_1;q_0]\bigr|_{q_1\to -q_1}$ and does not integrate over the paths between $q_0$ and $-q_1$. }

Hereafter, we will consider the DeWitt propagator based on the Lorentzian path integral. In the quantum cosmology framework, particularly the Picard-Lefschetz formulation of the \ac{bfv} path integral, it is not obvious whether the image method discussed in quantum mechanics applies to this. Moreover,
as shown in Appendix~\ref{sec:appendix1}, the image is only applicable for the symmetric potential with respect to the boundary, but the potential of the action~\eqref{Action} is not. Therefore, it is not easy to apply this method to the usual set-up where the positive spatial curvature $K=1$ and the cosmological constant. Thus, to avoid technical difficulty in this paper, let us consider a different cosmological setup with respect to the quantum creation of the universe.
Note that the model discussed below is identical to the infinite potential problem for free particles, but with Hamiltonian constraints since we are discussing quantum gravity.

First, let us consider the simplest case with zero cosmological constant 
$\lambda=0$, and the Wheeler-DeWitt equation reduces the following form,
\begin{equation}
\hat{\mathcal{H}} G[q]=\frac{N}{2}
\left(4 \frac{d^2}{d q^2}-1\right) G[q]=0 \,.
\end{equation}
Imposing the DeWitt boundary condition $G[q=0]=0$, we obtain 
the following solution, 
\begin{equation}\label{DW-Wheeler-DeWitt-zerocos}
G_{\rm DW}[q]=\mathcal{C}_{3} (e^{\frac{q}{2}}-e^{-\frac{q}{2}}).
\end{equation}
where $\mathcal{C}_{3}$ is a constant.

On the other hand, with zero cosmological constant 
$\lambda=0$ the DeWitt propagator~\eqref{DW-propagator} based on the Picard-Lefschetz formulation of the \ac{bfv} path integral is given by,
\begin{align}\label{DWPL-integral-zerocos}
G_{\rm DW}[q_{1};q_{0}] & =  \sqrt{\frac{i}{8\pi}}
\int_{-\infty}^{\infty} 
\frac{\mathrm{d}N}{N^{1/2}}
e^{i S_{\textrm{on-shell}}[q_{1},q_{0},N]}
-G[q_1;q_0]\biggr|_{q_1\to -q_1}\,,
\end{align}
where we integrate the lapse function in $(-\infty,\infty)$ to get the wave function and 
$S_{\textrm{on-shell}}[q_{1},q_{0},N]$ is written as,
\begin{align}
S_{\textrm{on-shell}}[q_{1},q_{0},N]= \frac{N}{2}\left(1-\frac{(q_{0}-q_{1})^2}{4 N^2}\right).
\end{align}
The lapse integral~\eqref{DWPL-integral-zerocos} can be estimated based on 
the two saddle points $N_s$,
\begin{equation}\label{saddle-zerocos}
N_s=\frac{ic_3}{2}(q_{0}-q_{1}),
\end{equation}
where $c_3\in\{-1,1\}$. According to the Picard-Lefschetz theory, the Lefschetz thimbles ${\cal J}_\sigma$ can only take one saddle point with $c_3=-1$ in the integration for $q_0 < q_1$. By integrating $N$ along the Lefschetz thimbles ${\cal J}_\sigma$, and using the saddle point approximation,
we can obtain,
\begin{align}\label{wavefunction-saddle}
G[q_{1};q_{0}]  & =  \sum_\sigma n_\sigma \sqrt{\frac{i}{8\pi}}
\int_{{\cal J}_\sigma} \frac{\mathrm{d}N}{N^{1/2}}
e^{i S_{\textrm{on-shell}}[q_{1},q_{0},N]} \nonumber\\ 
& \approx \sum_\sigma n_\sigma\sqrt{\frac{i}{8\pi}} \frac{e^{i S_{\textrm{on-shell}}^{\, \textrm{saddle}}}}{N_s^{1/2}}\int_{{\cal J}_\sigma} \mathrm{d}N \exp \left( {\frac{i}{2}\frac{\partial^2S_{\textrm{on-shell}}^{\, \textrm{saddle}}}{\partial N^2}\Bigr|_{N=N_s}
(N-N_s)^2} \right) \nonumber\\ 
& \approx \sum_\sigma n_\sigma\sqrt{\frac{i}{8\pi}} \frac{e^{i 
S_{\textrm{on-shell}}^{\, \textrm{saddle}}}}{N_s^{1/2}}e^{i\theta_\sigma} \int_{{\cal J}_\sigma} \mathrm{d}R \exp \left({-\frac{1}{2}\left|\frac{\partial^2S_{\textrm{on-shell}}^{\, \textrm{saddle}}}{\partial N^2}\right|R^2} \right)\nonumber\\ 
& \approx \sum_\sigma n_\sigma\sqrt{\frac{i}{4N_s 
\left|\frac{\partial^2S_{\textrm{on-shell}}^{\, \textrm{saddle}}}{\partial N^2}\right|}} e^{i\theta_\sigma} e^{i S_{\textrm{on-shell}}^{\, \textrm{saddle}}}\,,
\end{align}
where we expand the on-shell action around a saddle point and introduce 
$N-N_s \equiv R e^{i\theta_\sigma}$,
\begin{equation}
S_{\textrm{on-shell}} = S_{\textrm{on-shell}}^{\, \textrm{saddle}} + \frac{1}{2} \frac{\partial^2S_{\textrm{on-shell}}^{\, \textrm{saddle}}}{\partial N^2}\Bigr|_{N=N_s}(N-N_s)^2 + \dots\,, 
\end{equation}
and 
\begin{equation}
{\frac{i}{2}\frac{\partial^2S_{\textrm{on-shell}}^{\, \textrm{saddle}}}{\partial N^2}\Bigr|_{N=N_s}
(N-N_s)^2} =\frac{i}{2}
\left|\frac{\partial^2S_{\textrm{on-shell}}^{\, \textrm{saddle}}}{\partial N^2}\right|
R^2e^{i(2\theta_\sigma+\alpha)}=
-\frac{1}{2}\left|\frac{\partial^2S_{\textrm{on-shell}}^{\, \textrm{saddle}}}{\partial N^2}\right|R^2\,,
\end{equation}
where $\textrm{Arg}\left( \frac{\partial^2S_{\textrm{on-shell}}^{\, \textrm{saddle}}}{\partial N^2}\Bigr|_{N=N_s}\right)=\alpha$ and $e^{i(2\theta_\sigma+\alpha)}=i$, $\theta_\sigma=\pi/4-\alpha/2$.
By imposing the saddle point $N_s$~\eqref{saddle-zerocos} we obtain the following expression, 
\begin{align}
G[q_{1};q_{0}] & \simeq \frac{i}{2}e^{\frac{q_{0}}{2}}e^{-\frac{q_{1}}{2}}. 
\end{align}
Now we can get the DeWitt wave function using Eq.~\eqref{DWPL-integral-zerocos}, 
\begin{align}\label{wavefunction-saddle-zerocos}
G_{\rm DW}[q_{1};q_{0}] &= \sqrt{\frac{i}{8\pi}}
\int_{-\infty}^{\infty} 
\frac{\mathrm{d}N}{N^{1/2}}
e^{i S_{\textrm{on-shell}}[q_{1},q_{0},N]}
-G[q_1;q_0]\biggr|_{q_1\to -q_1}\nonumber \\
&\simeq \frac{i}{2}e^{\frac{q_{0}}{2}}
[e^{-\frac{q_{1}}{2}}-
e^{\frac{q_{1}}{2}}].
\end{align}
This result is consistent with the solution~\eqref{DW-Wheeler-DeWitt-zerocos} in the Wheeler-DeWitt equation.
The physical interpretation of~\eqref{wavefunction-saddle-zerocos} is that the wave function of a closed universe with 
absolutely no matter, satisfying the DeWitt boundary condition, predicts a universe where the probability at the initial singularity is zero, but sufficiently large for the large scale factor. However, the corresponding saddle points are purely imaginary, and 
such a universe obeying the DeWitt wave function does not satisfy the classical equations of motion and the Hamiltonian constraint, indicating quantum behavior.

Most studies of quantum cosmology consider the closed universe with positive curvature $K=1$ because this provides the elegant picture of the quantum creation of the universe from nothing and the action is finite. However, we can consider the creation of flat or open universes with zero or negative spatial curvature $K=0, -1$. In the general case, the spatial volume in three dimensions is infinite in these cases. Still, if the space-time topology is non-trivial, the creation of a compact flat or open quantum universe of space-time is possible~\cite{Zeldovich:1984vk,Coule:1999wg, Linde:2004nz, Linde:2017pwt}. For the closed universe, 
the wave function either behaves exponentially suppressed or amplified. On the other hand, for a flat or open compact universe, the wave function does not exhibit such exponential behavior and might be favored for probability interpretation and perturbation problems.
Next, let us consider the quantum creation of the flat universe with the cosmological constant $\lambda$.

For simplicity, we assume that the volume factor of the compact flat universe is the same as the closed universe, and the action is written as 
\begin{align}\label{Action-flat}
S[q,N]= \frac{1}{2}\int_{t=0}^{t=1} N\mathrm{d}t 
\left( -\frac{\dot{q}^2}{4 N^2} - \lambda q \right) \,.
\end{align} 
For convince, we redefine the lapse function $N(t)\to N(t)q^{-1}$ to render the action
quadratic in $\mathfrak{q}=\frac{2}{3}q^{3/2}$, and can obtain the following action,
\begin{align}\label{Action-flat-red}
S[\mathfrak{q},N]= \frac{1}{2}\int_{t=0}^{t=1} N\mathrm{d}t 
\left( -\frac{\dot{\mathfrak{q}}^2}{4 N^2} - \lambda \right) \,.
\end{align} 
We note that such a redefinition involves subtle issues since it changes the path integral measure. However, since the evaluation of the path integral in this paper is based on a semi-classical analysis, we can neglect these issues.
In this set-up, the Wheeler-DeWitt equation reads,
\begin{equation}\label{eq:WdW-equation-flat}
\hat{\mathcal{H}} G[\mathfrak{q}]=\frac{N}{2}\left(4 \frac{d^2}{d \mathfrak{q}^2}+\lambda \right) G[\mathfrak{q}]=0 \,.
\end{equation}
By imposing the DeWitt boundary condition $G[\mathfrak{q}=0]=0$, we can get, 
\begin{equation}\label{DW-Wheeler-DeWitt-flat}
G_{\rm DW}[\mathfrak{q}]=\mathcal{C}_{4}\sin \left(\frac{\sqrt{\lambda }}{2} \mathfrak{q}\right)\, ,
\end{equation}
where $\mathcal{C}_{4}$ is a constant.

Next, we will consider the wave function based on the path integral formulation.
For a flat universe with the non-zero cosmological constant, the DeWitt wave function~\eqref{DW-propagator} based on the Picard-Lefschetz formulation of the \ac{bfv} path integral is,
\begin{align}\label{DWPL-integral-flat}
G_{\rm DW}[\mathfrak{q}_{1},\mathfrak{q}_{0}] & =  \sqrt{\frac{i}{8\pi}}
\int_{-\infty}^{\infty} 
\frac{\mathrm{d}N}{N^{1/2}}
e^{i S_{\textrm{on-shell}}[\mathfrak{q}_{1},\mathfrak{q}_{0},N]}
-G[\mathfrak{q}_{1},\mathfrak{q}_{0}]\biggr|_{\mathfrak{q}_{1}\to -\mathfrak{q}_{1}}\,,
\end{align}
where we integrate the lapse function in $(-\infty,\infty)$ and $S_{\textrm{on-shell}}[\mathfrak{q}_{1},\mathfrak{q}_{0},N]$ is written as,
\begin{align}
S_{\textrm{on-shell}}[\mathfrak{q}_{1},\mathfrak{q}_{0},N]= -\frac{N}{2} \left(\lambda +\frac{(\mathfrak{q}_{1}-\mathfrak{q}_{0})^2}{4 N^2}\right).
\end{align}
The lapse integral~\eqref{DWPL-integral-flat} can be estimated based on 
the two saddle points $N_s$,
\begin{equation}\label{saddle}
N_s=c_4\frac{\mathfrak{q}_{0}-\mathfrak{q}_{1}}{2 \sqrt{\lambda }},
\end{equation}
where $c_4\in\{-1,1\}$ and the saddle-point lives in real Lorentzian axis.
Thus, the Lefschetz thimbles ${\cal J}_\sigma$ can pass the two saddle points with $c_3=\pm 1$. 
By integrating $N$ in 
along the Lefschetz thimbles ${\cal J}_\sigma$ in the DeWitt wave function~\eqref{DWPL-integral-zerocos} and assuming $\mathfrak{q}_0 < \mathfrak{q}_1$, 
we obtain,
\begin{align}
G_{\rm DW}[q_{1};q_{0}] & \simeq  2\sqrt{\frac{i}{\lambda}}e^{i\frac{5}{4}\pi}
\sin \left(\frac{\sqrt{\lambda}}{2}q_1\right)
\cos \left(\frac{\sqrt{\lambda}}{2}q_0\right), 
\end{align}
where we took saddle-point approximation for the two saddle points $N_s$ with $c_3=\pm 1$. This result is consistent with the above solution~\eqref{DW-Wheeler-DeWitt-flat} in the Wheeler-DeWitt equation.
Thus, the image method is also correct in Lorentzian quantum cosmology using the Picard-Lefschetz formulation of the \ac{bfv} integral.
The DeWitt wave function for such a constant potential is a standing wave. By using $\lambda=\frac{2G}{9\pi}\Lambda$, we have the relation $2\pi/\sqrt{\lambda}=\sqrt{18\pi^3/G\Lambda}$, suggesting that the DeWitt wave function tends to vanish around the Hubble horizon, $1/H_{\Lambda}\equiv \sqrt{3/\Lambda}$ characterized the cosmological constant $\Lambda$ and the probability of the wave function also vanish around the Hubble horizon.

\section{Conclusions}
\label{sec:conclusion}
In this paper, we have discussed the DeWitt wave function in the path integral formulation of quantum gravity. First, because imposing boundary conditions on wave functions based on the path integral formulation is conceptually and technically unclear in quantum mechanics, the infinite potential models have problems in the path integral formulation. The simplest way to provide the propagator vanishing at the boundary is the image method proposed in Refs.~\cite{Janke:1979fv,1981AmJPh..49..843G}. In this paper, we have discussed the image method in path integrals for systems such as free particles with infinite potential, infinite well potential, and half-harmonic oscillator models. These have been discussed in the literature~\cite{1984PhLA..106..212S,1985NuPhB.257..799A,
1993PhRvA..48.3445N,1997JPhA...30.5993A,2011arXiv1112.3674D}, but we have clearly summarised these results. However, we have shown that the image method only applies to symmetric potentials with respect to the boundary. For asymmetric potentials, the restricted propagator of the image method does not give the correct wave function that satisfies the Schr\"{o}dinger equation.

Based on this consideration, we have applied the image method to the framework of quantum cosmology on the basis of the Picard-Lefschetz formulation of the \ac{bfv} path integral and defined the DeWitt propagator and DeWitt wave function based on the Lorentzian path integral formulation. In the case of symmetric potentials written with scale factors with respect to $a=0$, these coincide with solutions of the Wheeler-DeWitt equation and predict a singularity-free creation of the universe. Unfortunately, in common quantum cosmology models, the potential of the scale factor is not symmetric with respect to $a=0$ due to the positive spatial curvature and the cosmological constant, and the imaging method does not work. We leave this difficulty as a future work and focus on different models of quantum cosmology where the superpotential is symmetric. We also note that changing the potential to be symmetric may solve the problem for the closed universe. It remains unclear whether this method remains effective when increasing the degrees of freedom, such as adding perturbations in path integrals. We plan to address these issues in future research.

Moreover, there are several boundary conditions applicable to the wave function of the universe, specifically, Dirichlet, Neumann, and Robin conditions. In this paper, we have adopted the simplest, DeWitt (Dirichlet) boundary condition. The applicability of the image method to other boundary conditions beyond the DeWitt boundary condition
is unclear since it only eliminates singular paths in the forbidden region. On the other hand, the path integral methods have been studied in quantum mechanics for various boundary conditions (see, e.g., \cite{Farhi:1989jz,Carreau:1990wh,PhysRevA.51.1811}). Extending these methods to quantum cosmology presents an intriguing prospect~\cite{Lemos:1995qu},  but we leave this to future work.

\section*{Acknowledgment}
H.M. would like to thank Shinji Mukohyama, Atsushi Naruko, and
Takahiro Terada for helpful discussions. 
H.M. would like to thank Ding Jia for useful comments. 
This work is supported by JSPS KAKENHI Grant No. JP22KJ1782 and No. JP23K13100.

\appendix

\section{Restricted propagator and 
Schr\"{o}dinger equation}
\label{sec:appendix1}
In this Appendix, we will show that for a potential symmetric with respect to the boundary, the restricted propagator $K_R\left[x', t'; x,t\right]$ in quantum mechanics derives a correct wave function that satisfies the Schr\"{o}dinger equation. 
We will adopt the argument often used to show the equivalence of the path integral and the Schrodinger equation~\cite{feynman2010quantum}, where we consider a system evolving from $(y,t)$ to $(x,t+\epsilon)$.
In order to show the relation between the restricted propagator and 
the Schr\"{o}dinger equation, we will consider the wave function given by the propagator $K\left[-x, t+\epsilon; y,t\right]$, 
\begin{align}
\Psi\left[x, t+\epsilon\right]=A \int^{\infty}_{-\infty}
dy K\left[-x, t+\epsilon; y,t\right]\Psi\left[y, t\right],
\end{align}
where $A$ is a normalization constant.
Since the time evolution of this system only changes at minute time $\epsilon$, we only need to discuss one approximate path,
\begin{align}
\Psi\left[x, t+\epsilon\right]\simeq A \int^{\infty}_{-\infty}
dy \exp \left[\frac{i}{2\hbar}\frac{m(x+y)^2}{\epsilon}
-\frac{i}{\hbar}\epsilon V\left(\frac{-x+y}{2}\right)\right]\Psi\left[y, t\right],
\end{align}
where the action is approximated at the midpoint. The term $\frac{m(x+y)^2}{\epsilon}$ appears in the first exponential part, and when $y$ is a finite distance from $-x$, $\frac{m(x+y)^2}{\epsilon}$ becomes very large in a certain limit, so the exponential factor oscillates very rapidly with respect to $y$ variation. When this factor oscillates rapidly, the main contribution is obtained only when $y$ is close to $-x$. For this reason, if we set $y=-x+\epsilon^{1/2}\eta$, we can expect only the smallest terms to give the main contribution to the integral,
\begin{align}
\Psi\left[x, t+\epsilon\right]\simeq \epsilon^{1/2}A \int^{\infty}_{-\infty}
d\eta \exp \left[\frac{im\eta^2}{2\hbar}
-\frac{i}{\hbar}\epsilon V\left(-x+\frac{\epsilon^{1/2}\eta}{2}\right)\right]\Psi\left[-x+\epsilon^{1/2}\eta, t\right].
\end{align}
By expanding each terms in terms of $\epsilon^{1/2}$ and taking into account up to the first order of $\epsilon$, we can obtain, 
\begin{align}
\begin{split}
\Psi\left[x, t\right]+\epsilon \frac{\partial }{\partial t}\Psi\left[x, t\right] &\simeq \epsilon^{1/2}A \int^{\infty}_{-\infty}
d\eta \exp \left(\frac{im\eta^2}{2\hbar}
\right)\Biggl[\Psi\left[-x, t\right]+\epsilon^{1/2}\eta \frac{\partial }{\partial x}\Psi\left[-x, t\right] \\
&+\frac{1}{2}\epsilon\eta^2 \frac{\partial^2 }{\partial x^2}\Psi\left[-x, t\right]-\frac{i}{\hbar}\epsilon V\left(-x\right)\Psi\left[-x, t\right]\Biggr].
\end{split}
\end{align}

Now, we will suppose that the potential is symmetry with respect to the boundary, i.e. $V\left(x\right)=V\left(-x\right)$ so that the wave function is also symmetric,
$\Psi\left[x, t\right]=\Psi\left[-x, t\right]$. In this case, we can compare the first term of the left-hand side and right-hand side of the above equation, and define the 
normalization constant $A$, 
\begin{equation}
\frac{1}{A}=\epsilon^{1/2}\int^{\infty}_{-\infty}
d\eta \exp \left(\frac{im\eta^2}{2\hbar}
\right)=\left(\frac{2\pi i\epsilon\hbar}{m}\right)^{1/2}.
\end{equation}
By using the above expression of $A$ we can get the Schr\"{o}dinger equation,
\begin{equation}
\left[
-\frac{\hbar^2}{2m}\frac{\partial^2}{\partial x^2}+V(x)-
i\hbar \frac{\partial}{\partial t}\right]\Psi\left[x, t\right]=0\,.
\end{equation}
The wave function derived by the propagator $K\left[-x', t'; x,t\right]$ satisfies the Schr\"{o}dinger equation, and thus, the wave function derived by the restricted propagator $K_R\left[x', t'; x,t\right]$ also satisfies the Schr\"{o}dinger equation from the principle of superposition as, 
\begin{align}
\begin{split}
\Psi\left[x', t'\right] &=\int\Psi\left[x, t\right]K_R\left[x', t'; x,t\right]\mathrm{d}x \\
 &=\int\Psi\left[x, t\right]K\left[x', t'; x,t\right] 
\mathrm{d}x -\int\Psi\left[x, t\right]K\left[-x', t'; x,t\right] 
\mathrm{d}x\;.
\end{split}
\end{align}

Another proof is directly given by the following argument; when we insert the wave function derived from the restricted propagator $K_R\left[x', t'; x,t\right]$ into the Schr\"{o}dinger equation, we can obtain, 
\begin{equation}
\left[
-\frac{\hbar^2}{2m}\frac{\partial^2}{\partial x'^2}+V(x')-
i\hbar \frac{\partial}{\partial t'}\right]\Psi\left[x', t'\right]=
-i\hbar\delta(t'-t)\Psi\left[x', t'\right]+
i\hbar\delta(t'-t)\Psi\left[-x', t'\right]=0\,,
\end{equation}
where we used $V\left(x\right)=V\left(-x\right)$ and 
$\Psi\left[x, t\right]=\Psi\left[-x, t\right]$.
Therefore, the image method is correct when the potential exhibits
symmetry with respect to the boundary.

\section{Lorentzian quantum cosmology and Picard-Lefschetz 
theory}
\label{sec:appendix2}

Hereafter, we shall review the basis of the Picard-Lefschetz formulation of the \ac{bfv} path integral~\cite{Feldbrugge:2017kzv} which allows the gravitational transition amplitude~\eqref{G-amplitude} to be carried out in the regular way.
First, we introduce the action for the convenience,
\begin{align}\label{Action}
S[q,N]= \frac{1}{2}\int_{t=0}^{t=1} N\mathrm{d}t 
\left( -\frac{\dot{q}^2}{4 N^2} + 1 - \lambda q \right) \,.
\end{align} 
Since $S[q,N]$ is quadratic, the path integral~\eqref{G-amplitude} 
can be evaluated under the semi-classical approximation.
We assume the full solution $q(t) = q_s(t) + Q(t)$ where
$Q(t)$ is the Gaussian fluctuation around the solution of the equations of motion,
\begin{align}
\delta S[q]/\delta{q} & =  0 \ \Longrightarrow \ 
\frac{\ddot{q}}{2N^2}=\lambda\label{eq:xeqn}.
\end{align}
with the boundary condition $q(0)=q_{0}$ and $q(1)=q_{1}$.
By substituting the solution for the action~\eqref{Action} and 
integrating the path integral over $Q(t)$, 
we have the following expression,
\begin{equation}\label{G-amplitude3}
G[q_{1};q_{0}]  =  \sqrt{\frac{i}{8\pi}}\int \frac{\mathrm{d} N}{N^{1/2}} 
e^{i S_{\textrm{on-shell}}[q_{1},q_{0},N]},
\end{equation}
where $S_{\textrm{on-shell}}[q_{1},q_{0},N]$ is the on-shell action,
\begin{align}\label{on-shell}
S_{\textrm{on-shell}}[q_{1},q_{0},N]&=\frac{1}{2}\int_{0}^{1} N\mathrm{d}t 
\left( -\frac{\dot{q_s}^2}{4 N^2} + 1 - \lambda q_s\right) 
\nonumber \\
&= \frac{\lambda ^2 N^3}{24}-\frac{N}{4}\left(\lambda (q_0+q_1)-2\right)-\frac{(q_0-q_1)^2}{8 N}\,.
\end{align}

Now, we shall evaluate the gravitational transition amplitude~\eqref{G-amplitude3} using the Picard-Lefschetz theory.
The Picard-Lefschetz theory complexifies a variable and 
selects a complex integration contour
based on the steepest descent paths $\cal J_\sigma$ 
known as the Lefschetz thimbles. 
Hence, we integrate $N$ 
along the Lefschetz thimbles ${\cal J}_\sigma$ as,
\begin{align}\label{PL-integral}
G[q_{1};q_{0}] & =  \sum_\sigma n_\sigma \sqrt{\frac{i}{8\pi}}
\int_{{\cal J}_\sigma} \frac{\mathrm{d}N}{N^{1/2}}
e^{i S_{\textrm{on-shell}}[q_{1},q_{0},N]}.
\end{align}
The lapse integral~\eqref{PL-integral} can be estimated based on 
the four saddle points $N_s$,
\begin{equation}\label{saddle-lapse-L}
N_s=c_1\frac{1}{\lambda}
\left[\sqrt{q_{0}\lambda-1}
+c_2\sqrt{q_{1}\lambda-1}\right],
\end{equation}
where $c_1,c_2\in\{-1,1\}$.
The four-saddle points~\eqref{saddle-lapse-L} correspond to the intersection
of the Lefschetz thimble $\cal J_\sigma$ and steepest ascent 
paths $\cal K_\sigma$ 
where $\textrm{Re}\left[iS_{\textrm{on-shell}}\left( N \right)\right]$
decreases and increases monotonically on $\cal J_\sigma$ and $\cal K_\sigma$,
and $n_\sigma$ is the intersection number.
By imposing $q_0 = 0$~\cite{Hartle:1983ai}, 
and assuming $q_{1}\lambda > 1$,
the gravitational transition amplitude~\eqref{G-amplitude3} 
corresponds to the transition amplitude created from nothing.
Based on the Lorentzian Picard-Lefschetz method,
Feldbrugge $\textrm{et al.}$~\cite{Feldbrugge:2017kzv} showed that 
the gravitational transition amplitude~\eqref{G-amplitude3} 
by perfuming the integral over a contour in $(0,\infty)$
reduces to the tunneling propagator.
On the other hand, by integrating the lapse over contours in $(-\infty,\infty)$
where the integration of \eqref{G-amplitude3} has a singularity at $N = 0$, 
the choice of the path above or below the singularity at $N = 0$ provides the tunneling wave function~\cite{Feldbrugge:2017mbc} or the no-boundary wave function~\cite{DiazDorronsoro:2017hti}. 
Although $N>0$ is only allowed to distinguish $q_{1}$ from $q_{0}$ by the causality constraints~\cite{Teitelboim:1981ua,Teitelboim:1983fh}, we shall consider all integration of $N$ to derive the cosmological wave function.

Strictly, the no-boundary saddle points require a change of the Lefschetz thimbles in the complex lapse plane. This may be somehow problematic from the view of the Picard-Lefschetz theory and we only consider the integration of the path above the singularity at $N = 0$. By integrating the lapse over contours in $(-\infty,\infty)$ the 
the gravitational transition amplitude~\eqref{G-amplitude3} with $q_{0}=0$ reduces the tunneling wave function, 
\begin{align}
\begin{split}\label{eq:tunneling-wave-function}
G[q_{1};0] &\simeq \mathcal{A}_1\, e^{\frac{-1-i (\lambda q_{1}-1)
\sqrt{\lambda  q_{1}-1}}{3 \lambda }}+
\mathcal{A}_1^{\dagger}\, e^{\frac{-1+i (\lambda q_{1}-1) \sqrt{\lambda q_{1}-1}}{3 \lambda }}\\
&=\left|\mathcal{A}_1 \right|e^{-\frac{1}{3 \lambda }}\cos
\left(-\frac{(\lambda q_{1}-1)^{\frac{3}{2}}}{3 \lambda }+\rm{Arg[\mathcal{A}_1]}\right),
\end{split}
\end{align}
where $\mathcal{A}_{1}$ are prefactor and the contribution of each 
tunneling saddle points is a complex conjugate~\cite{Feldbrugge:2017mbc}.
In the Wheeler-DeWitt equation~\eqref{eq:WdW-equation}, the general solution 
is given by 
\begin{align}
\Psi\left[q_{1}\right] =\mathcal{C}_1\mathrm{Ai}\left(\frac{1-\lambda q_{1}}{(-2 \lambda)^{2 / 3}}\right)+\mathcal{C}_2\mathrm{Bi}\left(\frac{1-\lambda q_{1}}{(-2 \lambda)^{2 / 3}}\right)  \;,
\end{align}
where $\mathrm{Ai}(z), \mathrm{Bi}(z)$ are Airy functions of the first and second kind, and $\mathcal{C}_{1,2}$ are constants. In the asymptotic form of Airy function at the limit $q_{1}\to 0$ we have
\begin{align}
\mathrm{Ai}\left(\frac{1-\lambda q_{1}}{(-2 \lambda)^{2 / 3}}\right)
\sim \frac{1}{\sqrt{\pi}\left(\frac{1-\lambda q_{1}}{(-2 \lambda)^{2 / 3}}\right)^{\frac{1}{4}}}\cos\left(-\frac{(\lambda q_{1}-1)^{\frac{3}{2}}}{3 \lambda }-\frac{1}{4}\pi \right),
\end{align}
which suggests the wave function~\eqref{eq:tunneling-wave-function} is consistent with the above solutions of the Wheeler-DeWitt equation~\eqref{eq:WdW-equation}.
The rigorous proof is given in the literature~\cite{Feldbrugge:2017kzv,Feldbrugge:2017mbc}.

\nocite{}
\bibliography{reference}
\bibliographystyle{JHEP}
\end{document}